\documentclass[useAMS,usenatbib]{mn2e}

\usepackage{times}

\usepackage{graphicx}
\usepackage{amsmath}
\usepackage{amssymb}
\usepackage{natbib}

\newcommand{\apj}{Ap.J.}
\newcommand{\aj}{A.J.}
\newcommand{\mnras}{MNRAS}
\newcommand{\aap}{A\&A}
\newcommand{\pasp}{PASP}
\def\trh0{t_{rh}(0)}

\def\apgt{\ {\raise-.5ex\hbox{$\buildrel>\over\sim$}}\ }
\def\aplt{\ {\raise-.5ex\hbox{$\buildrel<\over\sim$}}\ }
\def\ltorder{\aplt}
\def\gtorder{\apgt}

\title[Star Cluster Evolution with Primordial Binaries I]
{Star Clusters with Primordial Binaries:\break
I. Dynamical Evolution of Isolated Models}
\author[D.C. Heggie, M. Trenti and P. Hut]
{Douglas C. Heggie$^{1}$\thanks{E-mail addresses:
d.c.heggie@ed.ac.uk (DCH); m.trenti@sns.it (MT); piet@ias.edu (PH)},
Michele Trenti$^{2,3}$\footnotemark[1] and Piet Hut$^{4}$\footnotemark[1]\\
$^{1}$School of Mathematics,
        University of Edinburgh,
        King's Buildings, Edinburgh EH9 3JZ,
        Scotland, U.K.\\
$^{2}$Scuola Normale Superiore, Piazza dei Cavalieri 7,
        I-56126 Pisa, Italy\\
$^{3}$Yukawa Institute for Theoretical Physics, Kyoto University,
606-8502 Kyoto, Japan\\
$^{4}$Institute for Advanced Study, Princeton, NJ 08540, USA}
\begin{document}

\date{Accepted ; Received ; in original form }

\pagerange{\pageref{firstpage}--\pageref{lastpage}} \pubyear{2005}

\maketitle

\label{firstpage}

\begin{abstract}
  In order to interpret the results of complex realistic star cluster
  simulations, which rely on many simplifying approximations and
  assumptions, it is essential to study the behavior of even more
  idealized models, which can highlight the essential physical effects
  and are amenable to more exact methods.  With this aim, we present
  the results of $N$-body calculations of the evolution of equal-mass
  models, starting with primordial binary fractions of 0 - 100 \%,
  with values of $N$ ranging from 256 to 16384.  This allows us to
  extrapolate the main features of the evolution to systems comparable
  in particle number with globular clusters.
  
  In this range, we find that the steady-state `deuterium main
  sequence' is characterized by a ratio of the core radius to
  half-mass radius that follows qualitatively the analytical estimate
  by \citet{vc94}, although the $N$ dependence is steeper than 
  expected. Interestingly, for an initial binary fraction $f$ greater than
  $10\%$, the binary heating in the core during the post collapse phase
  almost saturates (becoming nearly independent of $f$), and so little
  variation in the structural properties
is observed. Thus, although we observe a significantly lower binary
  abundance in the core with respect to the Fokker-Planck simulations
  by \citet{gao91}, this is of  little dynamical consequence. 
  
  At variance with the study of \citet{gao91}, we see no sign of
  gravothermal oscillations before $150$ halfmass relaxation times. At
  later times, however, oscillations become prominent. We demonstrate
  the gravothermal nature of these oscillations.

\end{abstract}


\section{Introduction}

The evidence for primordial binary stars in old globular star clusters
continues to accumulate steadily.  Even though no comprehensive review
has appeared since  \citet{hut92}, recent additions include
\citet{pul03}, \citet{bel02}, \citet{alb01}.  

The $N$-body modeling of such clusters has also developed rapidly,
with studies devoted to exotic species associated with binaries, blue
stragglers, intermediate mass black holes, hierarchical systems, and
so on \citep[see, for example, ][]{hur01,por02,aar01,eco04,hur05}.

There have been relatively few $N$-body studies devoted to the effects
of binaries on the structural evolution of model clusters.
\citet{mcm90}, \citet{mcm91}, \citet{mcm94}, \citet{heg92},
\citet{aar93}, \citet{kro95} and \citet{del96} described a wide range
of results with values of $N$ which are generally rather modest by
modern standards (though not inappropriate for application to many
open clusters).  More recent studies include that of \citet{wil03},
whose focus was on the evolution of the core radius with age.
Generally speaking, however, in recent years the effect of primordial
binaries on the structural evolution of rich systems has been studied
using approximate methods based on the Fokker-Planck equation or
related approaches.  The study by \citet{gao91} [hereafter Gao et
al.], using the Fokker-Planck model with finite differences,
established an interesting set of parameters which has been taken up
by \citet{gie00,gie03}, using their hybrid gas/Monte Carlo code, and
by \citet{fre03,fre05}, whose code was pure Monte Carlo {(with computed
binary interactions in the later paper)}.

For all the usual reasons, it is advisable to check the results of
approximate methods by means of $N$-body simulations, and it is the
main purpose of the present paper to do so, based on the standard
model defined by Gao et al.  We present the results of an extensive
set of direct $N$-body simulations, performed with up to $16384$
particles and different ratios for the number of initial binaries to
single stars, ranging from $0\%$ to $100\%$.  Following an outline of
the physical picture in Sec.~\ref{sec:physics}, the specification of
our simulations is given in Sec.~\ref{sec:simulations}.

The subsequent two sections contain the main presentation of our
results, with emphasis on (i) a comparison with the results of Gao et
al. (Sec.~\ref{sec:gao}), and (ii) an analysis in terms of the dependence
on $N$ and on the primordial binary fraction (Sec.~\ref{sec:ndependence}).
In fact these topics cannot be cleanly separated.  For example, the
main systematic differences observed with respect to Gao et al.  are
in the fraction of binaries in the core, and possibly in the size of
the core, in the post-collapse phase.  After extrapolating our results
to the number of particles used by Gao et al., we clearly observe a
lower fraction of binaries in the core (by around a factor $3$).
Extrapolation of the core radius is more problematic, however (see
Sec.~\ref{sec:rc}).  In this section we also compare our results with
those of other approximate numerical methods, and with the analytical
estimate of the core radius due to \citet{vc94}.  The last section of
the paper provides an extended summary.

\section{The Physical Picture}\label{sec:physics}

\begin{figure}
\resizebox{\hsize}{!}{\includegraphics{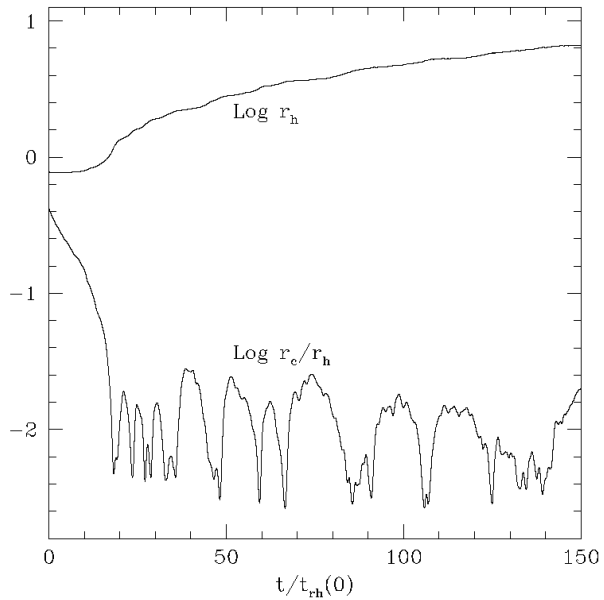}}
\caption{Dependence of half-mass (upper curve) radius (N-body units)
and of the ratio $r_c/r_h$ on time (units of the initial half-mass
relaxation time). The simulation has been performed with $16384$
particles and no primordial binaries.}\label{fig:1}
\end{figure}

For an isolated cluster, the most important characteristics describing
the mass distribution are the core radius $r_c$ and the half-mass
radius $r_h$.  Fig.~\ref{fig:1} shows how both evolve for the 
case of a Plummer model distribution of single stars, all of which
have equal mass, with no primordial binaries present.  The half-mass
radius remains roughly constant during the core collapse phase, while
the core shrinks at an increasing rate until the first core bounce
takes place.  Around this time, one or more binaries are formed
dynamically, through simultaneous three-body encounters.  The energy
production in those binaries halts core collapse, and also initiates
an expansion of the
half-mass radius.  In the post-collapse phase, the core
expands and contracts dramatically, while new binaries are formed and
older binaries are first hardened and then ejected.

\begin{figure*}
{\includegraphics[width=14.8cm]{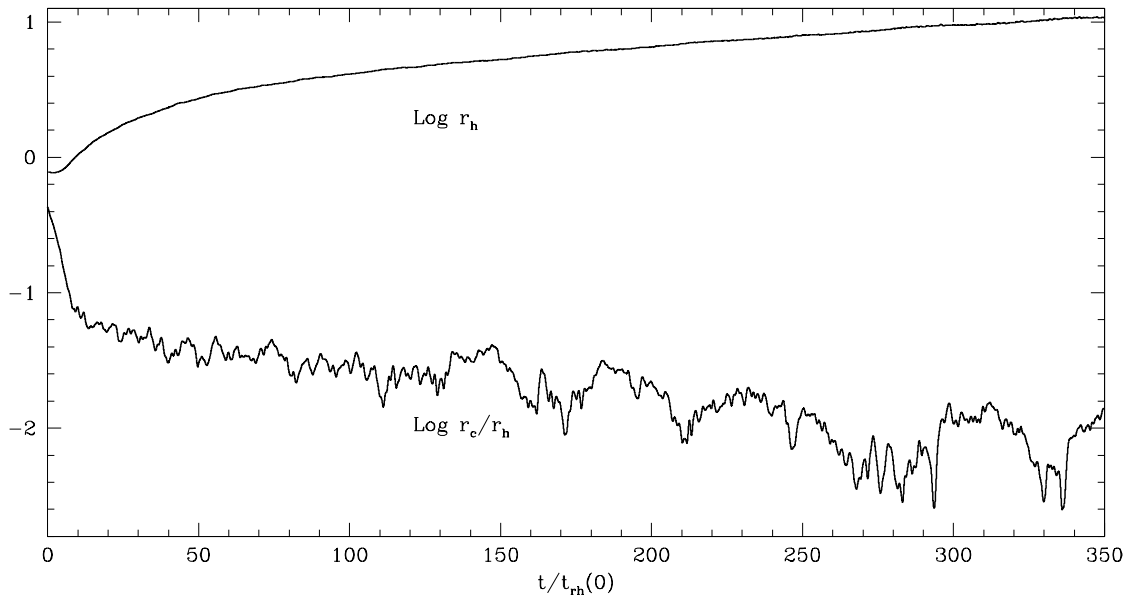}}
\caption{Dependence of half-mass (upper curve) radius (N-body units)
and of the ratio $r_c/r_h$ on time (units of the initial half-mass
relaxation time). The simulation has been performed with $16384$
particles and $10\%$ binaries (as defined in Eq.~\ref{eq:f}).}\label{fig:2}
\end{figure*}

Fig.~\ref{fig:2} shows the striking difference that primordial
binaries make to this simple picture. The presence of primordial
binaries causes the core to shrink roughly twice as fast as in the
single-star case.  This effect is largely due to mass segregation: the
binaries, between twice as massive as single stars, tend to
precipitate toward the center.  In contrast, the core contraction is
much less deep in the case of primordial binaries: the core
contraction levels off after the core shrinks by less than a factor
ten, whereas in the single star case core collapse continues till the
core is one hundredth of its original size.  Another difference is
that in the primordial binary case, the half-mass radius $r_h$ starts
increasing very early on, a factor four earlier than in the single
star case.  Since there is very little mass loss in these early stages
of cluster evolution, an increase in half-mass radius indicates an
injection of energy into the bulk motion of the stars and the center
of masses of the binaries.  This points to an efficient energy
mechanism at work, which then also explains the early halting of the
core contraction.

This is indeed the physical explanation.  Binaries can give off energy
in encounters between binaries and single stars\footnote{Though such encounters can result in the ejection of single stars and binaries, the cluster still gains energy, by the process known as ``indirect heating''.} (and also in
binary-binary encounters), the probability of which is given by the
product of the densities $\rho_b\rho_s$ of binaries and single stars,
whereas the formation rate of new binaries is proportional to
$\rho_s^3$.  The relevant scales for these densities are set by the
ninety degree turn-around distance $d$, which is of order $1/N$ in
standard units, far smaller than typical inter-particle distances, even
in the core.  If we define the interaction densities as the number of
particles per unit of volume $d^3$, and denote them by a hat, we have
$\hat\rho_s\ll 1$ and $\hat\rho_b\ll 1$.  In a model with primordial
binaries, $\hat\rho_s$ and $\hat\rho_b$ are comparable, at least in
order of magnitude, and it follows that $\hat\rho_b\hat\rho_s \gg
\hat\rho_s^3$.  In other words, at a given density the rate of energy
generation by primordial binaries far exceeds the rate of energy generation by
binaries that first have to be formed at that density \citep{hut89}.

Another way of describing the difference between these two forms of
evolution is that Fig.~\ref{fig:1} shows a collapse to the `hydrogen
main sequence', whereas Fig.~\ref{fig:2} shows a collapse to the
`deuterium main sequence', to borrow metaphors from stellar evolution.
Let us take a gas cloud of a solar mass, consisting of only hydrogen
and helium.  When this protostar contracts sufficiently, nuclear
reactions in the core will at some point balance the radiation losses
at the photosphere.  At this point, by definition the star has landed
on the (hydrogen) main sequence.  If the star would be made of pure
helium, it would shrink further, until it would reach the helium main
sequence, where the central temperature and pressure are high enough
to generate the same energy at the rate at which it is lost at the
surface.  In practice, however, protostars never contain pure
hydrogen, but typical have a very small admixture of deuterium.  This
deuterium can burn to form helium at lower densities and temperatures
than hydrogen can, for a similar reason that primordial binaries can
`burn' gravitationally at lower densities than single stars. As a
result, a contracting protostar first gets hung up at the deuterium
main sequence, where it has a radius significantly larger than it will
have soon afterwards, when its central deuterium reservoir is burned
up, and the star settles on the (hydrogen) main sequence.

Here we witness something similar: our contracting cluster core
settles onto an almost stable extended core configuration, while
burning the primordial binaries.  Gradually, when this primordial fuel
is being exhausted, the core shrinks since the burning process becomes
less efficient when the binary fraction becomes less.  This slow
shrinking of the core continues until (almost) all primordial binaries
are burned up, after which the new binaries have to be formed by the
process of simultaneous three-body encounters between single stars.
Note that the $r_c$ value in Fig.~\ref{fig:2} toward the end is
dropping toward the values seen in Fig.~\ref{fig:1} from the moment of
the first core collapse.  This behavior is more evident in
Fig.\ref{fig:gravo}. 

\begin{figure}
\resizebox{\hsize}{!}{\includegraphics{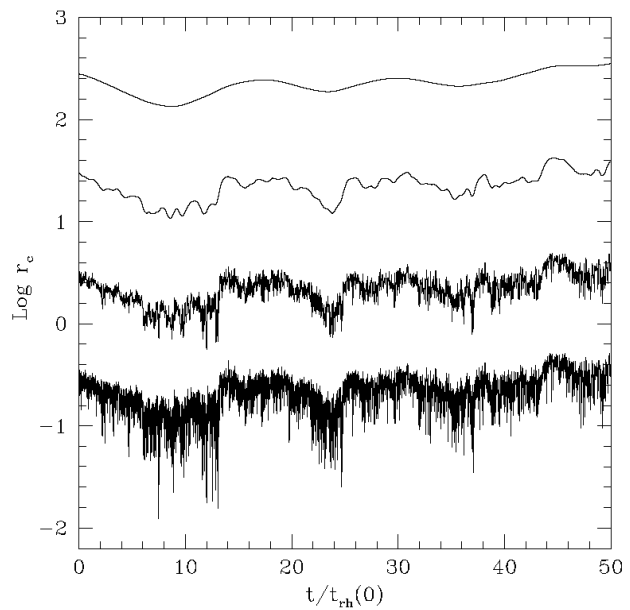}}
\caption{Evolution of the core radius versus time for different
  amounts of smoothing. From top to bottom the length of the smoothing
  window is $10 t_{rh}(0)$,$1 t_{rh}(0)$, $1 t_{d}$, $10^{-2}
  t_{d}$. For all but the last one, the curves are shifted upward by a
  constant factor for a better graphical representation.  The
  simulation has been performed with $1024$ particles and $10\%$ 
  binaries.}\label{fig:smooth}
\end{figure}

\section{Simulations: Setup and Analysis}\label{sec:simulations}

\subsection{Initial conditions and definitions}

The models shown in Figs.~\ref{fig:1} and \ref{fig:2}, like all those
studied in this paper, are isolated, with stars of equal mass.  The
initial distribution is a Plummer model.  In the standard model of Gao
et al., initially 10\% of the objects are binaries, as in
Fig.~\ref{fig:2}, with energies in the range $3~kT_c$--$400~kT_c$,
where $T_c$ is the central temperature of the system. For a Plummer
model $T_c \approx 1.7~T$, where $(3/2)kT$ is the mean kinetic energy
per particle of the system (the binaries being replaced by their
barycenter).  Gao et al. used Fokker-Planck models with parameters
corresponding to $N=3~10^5$ particles to study the evolution.  They
also carried out runs with the initial fraction of binaries equal to
$0.05$ and $0.2$.

In this paper we report on the result of $N$-body calculations, which
we performed using Aarseth's NBODY6 code \citep{aar03}, which has
been slightly modified to provide additional diagnostics on the
spatial distributions of single and binary stars (see, e.g.,
Fig.~\ref{fig:3a}).  Starting from a minimum of 256 particles, we have
made runs with no primordial binaries, 2\% primordial binaries, the
three Gao et al. choices of 5\%, 10\%, and 20\% primordial binaries,
as well as runs with 50\% and 100\% primordial binaries (Table
\ref{tab:tb1}).  As in the case of Gao et al., we have defined here
the primordial binary fraction as the fraction of objects that are binaries:
\begin{equation}\label{eq:f}
f = N_b/(N_s+N_b)
\end{equation}
where $N_b$ and $N_s$ are the initial number of binaries and single stars,
respectively.  This implies that the fraction of the total mass in
binary stars, in the case of equal masses, is larger in the following
way:
$$
f_m = 2N_b/(N_s+2N_b)
$$
For example, for a run with 10\% primordial binaries, $f=0.1$ whereas
$f_m=2/11\sim0.18$.

Note that  $N$ denotes the number of original objects, i.e.
$N=N_s+N_b$.  When we discuss a run with $N=1024$ and 50\% primordial
binaries, we are dealing with $1536$ stars.

All our results are presented using standard units \citep{hm86}
in which
$$
G = M = - 4 E_{\rm T} = 1
$$
where $G$ is the gravitational constant, $M$ the total mass, and
$E_{\rm T}$ the total energy of the system of bound objects.  In
other words, $E_{\rm T}$ does not include the internal binding
energy of the binaries, only the kinetic energy of their
center-of-mass motion and the potential energy contribution where each
binary is considered to be a point mass.  The corresponding
unit of time is $t_d = (GM)^{5/2}/(-4E_{\rm T})^{3/2}=1$.   
For the relaxation time, we use the following expression
$$
t_{rh} = \frac{0.138 N r_h^{3/2}}{\ln{(0.11 N)}}.
$$

A quantity of great interest is the core radius, which we here define
as in NBODY6, i.e.  essentially 
\begin{equation*}
 \label{eq:rc} {r}_c=
\sqrt{\frac{\sum_{i=1,\tilde{N}}{r_i^2
\rho_i^2}}{\sum_{i=1,\tilde{N}}{\rho_i^2}}}, 
\end{equation*}
 where $r_i$ is the
distance of the $i$th star from the density centre, and the density
$\rho_i$ around each particle is computed from the distance to the
fifth nearest neighbour \citep{cas85}.  In practice outlying stars are
omitted from the sum.

The runs were performed using idle time on many workstations through
the Condor system\footnote{http://www.cs.wisc.edu/condor/}: this
allowed us to obtain in a few months what would have taken a few years
on a single dedicated workstation.

\begin{center}
\begin{table}
\caption{N-body models with Gao et al.-like initial conditions}
\label{tab:tb1}
\begin{tabular}{llllllll}
$N$~~~~~$f$: & 0\% & 2\% & 5\% & 10\% & 20\% & 50\% & 100\% \\
\hline
256  & 43 & 2 & 2& 46 & 39 & 16 & 2 \\
512  & 49 & 2 & 3& 46 & 45 & 24 & 1 \\
1024 & 51 & 3 & 4 & 45 & 44 & 46 & 1 \\
2048 & 47 & 3 & 2 & 36 & 31 & 21 & 2 \\
4096 & 9 & 3 & 2 & 13 & 8 & 8 & 2 \\
8192 & 1 & 1 & 1 & 3 & 1 &  &  \\
16384& 1 &  &  & 1 &  &  &  \\
\hline
\end{tabular}
\newline Note: This table gives the number of realisations for each 
value of $N$ and $f$.
\\ {\sl}
\end{table}
\end{center}

\subsection{Fluctuations}

One significant difference between the $N$-body models we are using
and the models of Gao et al is the presence, in $N$-body models, of
fluctuations.  These have both a cosmetic and a scientific importance,
and we consider these issues together here.

In presenting the results from $N$-body simulations, there is always
the question of how much smoothing to apply.  In Fig. 3, we present
the results for one run, for the change in time of the core radius,
for a variety of different amounts of smoothing.  In some figures
in this paper we apply smoothing in
order to highlight the particular points made in each figure.
Fig.~\ref{fig:smooth} can be used as a gauge by which to judge the
amount of `noise' present in various figures.  Generally we have used
a smoothing interval of $2.5\trh0$.

These fluctuations also have a scientific interest.  In the context of
three-body binaries (i.e. those formed in three-body encounters,
without a population of primordial binaries) their role was considered
in simplified models by \citet{1988fbp..coll..319I} and
\citet{1991PASJ...43..589T}.  The role of fluctuations in an $N$-body
model, though in the absence of binaries, was considered by
\citet{1985MNRAS.212..189B}.  They showed that values of the velocity
dispersion, sufficiently well defined to determine its spatial
gradient, could be determined in a 1000-body system, if averaged over
a shell containing at least 90 particles and over at least one
crossing time.  With our present models we have an opportunity to
consider similar issues for the generation of energy by primordial
binaries.

\begin{figure}
\resizebox{\hsize}{!}{\includegraphics{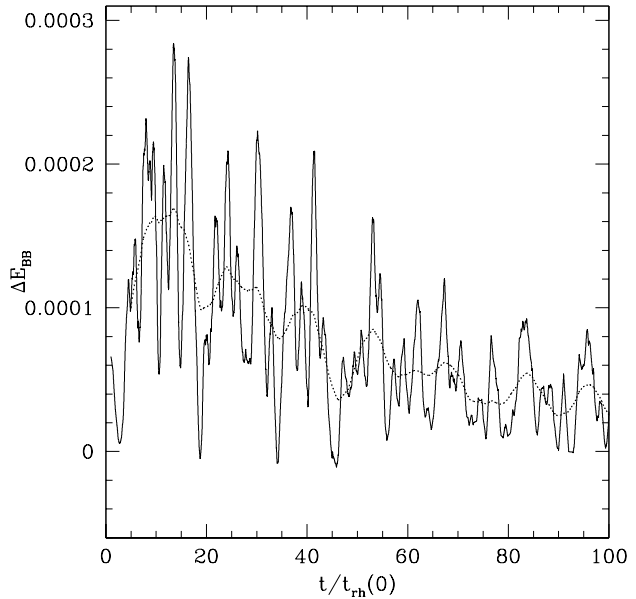}}
\caption{Change in internal energy of bound binaries, per unit time.
  The results have been smoothed with two triangular windows of full
  width $2~t_{rh}(0)$ (solid) and $10~t_{rh}(0)$
  (dotted).}\label{fig:energy_fluctuations}
\end{figure}

We restricted attention to our largest simulation ($N=16K, f=0.1$) and
considered the change in internal energy of all the bound binaries.
Within one $N$-body time unit this is usually zero, but may be
negative (if a binary hardens, for example) or positive (if a binary
escapes).  Perhaps surprisingly, it is necessary to smooth over a
long time interval (at least $2t_{rh}(0)$) in order for the result to
have a consistent sign (Fig.\ref{fig:energy_fluctuations}).

We have also considered how these fluctuations affect the rate of
interactions, in the following way.  If the proportion of binaries in
the core is very high, the rate of binary-binary interactions is
$\dot N_{bb}\propto \rho_c^2r_c^3$, where $\rho_c$ is the central
density.  Since $\rho_c r_c^2\propto v_c^2$ it follows that
$\dot N_{bb} \propto v_c^4/r_c$.  Therefore a simplified model, such as
a Fokker-Planck model, in which binaries interact at the correct
average rate, will yield a core with parameters $v_{FP}, r_{FP}$ such
that $\displaystyle{\frac{v^4_{FP}}{r_{FP}} =
  \langle\frac{v^4_{c}}{r_{c}}\rangle}$.  Fluctuations in an $N$-body
model, however, will lead to values of the mean core radius and
central velocity dispersion such that 
\begin{equation}\label{eq:inequation}
{\langle\frac{v^4_{c}}{r_{c}}\rangle\ne
\frac{\langle v_c^2\rangle^2}{\langle r_c\rangle}},
\end{equation}
and so one need
not expect these mean values to agree with those of a Fokker-Planck
model, even if the rate of energy generation by binaries in the
Fokker-Planck model is correct.  However, we have checked that the
difference in eq.(\ref{eq:inequation}), with smoothing as in
Fig.\ref{fig:energy_fluctuations}, is very small.  In a core supported
by primordial binaries, the core contains many particles, and so
fluctuations are unimportant by comparison, say, with a core supported
only by three-body binaries.

\section{Comparison with Gao et al.}\label{sec:gao}

The main purpose of this section is to present results of $N$-body
simulations which may be compared rather directly with those in Gao et
al. Therefore the sequence of presentation closely follows the
figures in that paper.  The main difference between our runs and those
of Gao et al. is in the particle number.  While Gao et al. used
parameters appropriate to $N= 3\times10^5$, even our largest run is
considerably smaller.  Our discussion of the $N$-dependence of our
results is mainly given later, in Sec.~\ref{sec:ndependence}.  The runs
discussed by Gao et al. have also been taken as a test case by
\citet{gie00} and by \citet{fre03,fre05}, and we also make occasional brief
comparisons with their results.

\subsection{Core and Half-Mass Radius}\label{sec:rcrh}

The structural evolution is best captured by the half-mass and core
radii. Fig.~\ref{fig:2} shows the results for a run with $N=8192$ and
10\% primordial binaries, and can be compared with Fig.~1 in Gao et
al., except that they scale radii by the initial scale radius of the
Plummer model (i.e. about 0.59 in our $N$-body units).  The expansion
of $r_h$ is similar (about 0.7 dex in both calculations), and the main
differences are in the evolution of the core radius.  These are
\begin{enumerate}
\item in the $N$-body model the initial decrease of $r_c$ is smaller,
  and $r_c$ remains considerably larger than in the Fokker-Planck
  model; and
\item in the $N$-body models (even for $N=16384$), there is no {
  sudden onset of
  deep core collapses, as observed in the Fokker-Planck model at
  about $t/t_{rh}(0)\simeq50$.}
\end{enumerate}
We now discuss these two issues in turn.

The fact that the contraction of the core is less deep than in the
model of Gao et al. could partly be an effect of the $N$-dependence
which we observe in our $N$-body models (Fig.~\ref{fig:rcrh}).  Our
conclusion, however, depends on how the extrapolation to $N =
3\times10^5$ (the value used by Gao et al.) is carried out.  If we
adopted the theory of \citealt{vc94} (see Equation~\ref{eq:vc}) we
would expect $r_c/r_h\simeq0.04$, which still exceeds by about 60\%
the value of about 0.025 found by Gao et al.  The data from our
simulations, however, suggest a steeper decrease of the core size with
$N$, and so the number given by Gao et al. could well be consistent
with our direct simulations.

The difference in values of $r_c$ could be caused by other factors,
including the relative abundance of single and binary stars
(Fig.~\ref{fig:3b}), which is also $N$-dependent
(Fig.~\ref{fig:nbns}).  A further factor to be borne in mind is that
the core radius should depend on the efficiency at which binaries
produce energy in three- and four-body collisions, and Gao et al. were
well aware of the uncertainties in the analytical cross sections
which they adopted.  Finally, the definition of core radius in the two
types of simulation may be different, though the initial values are
quite similar. 

Comparison with other results in the literature is beset by similar uncertainties.  We report simply that \citet{fre03} found an initial
contraction of the core fairly comparable to our result ({\sl despite}
the different values of $N$), {but then \citet{fre05} obtained a result
  comparable with that of \citet{gao91} when binary interactions were
  computed directly.}   In the hybrid model of
\citet{gie00}, on the other hand, the initial decrease of the core radius was only about
0.2 dex (Table \ref{tab:tb2}).

\begin{figure}
\resizebox{\hsize}{!}{\includegraphics{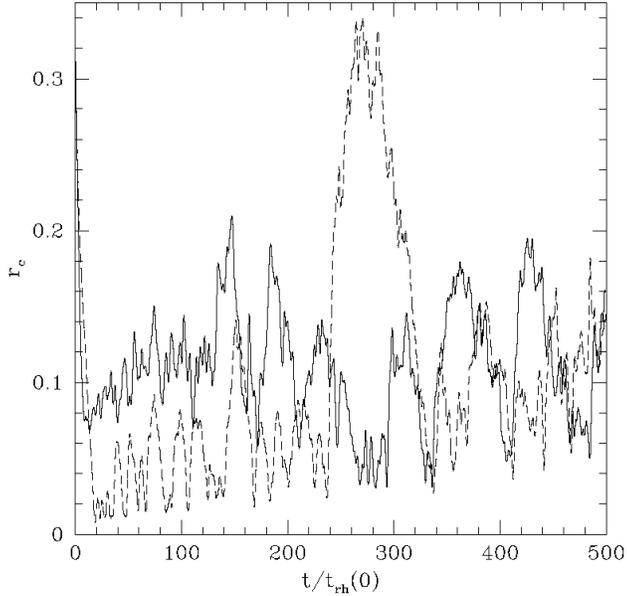}}
\caption{Gravothermal oscillation in the runs with $N=16384$ and 10\%
  primordial binaries run (continuous curve) and no primordial
  binaries (dashed line).  The data have been smoothed with a
  smoothing interval 2.5 $t_{rh}(0)$.}\label{fig:gravo}
\end{figure}

\begin{figure}
\resizebox{\hsize}{!}{\includegraphics{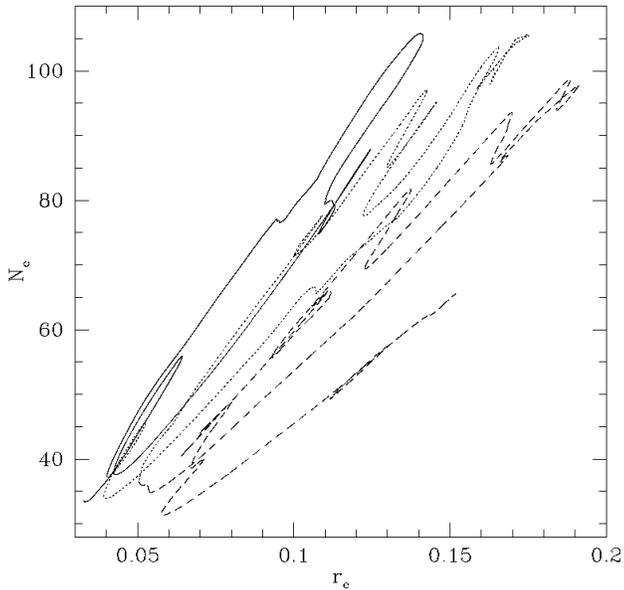}}
\caption{Three successive clockwise gravothermal oscillations in the
  $N=16384$ 10\% primordial binaries run.  The number of particles in
  the core is plotted against the core radius.  An isothermal core is
  a two-parameter family, and the presence of loops in such a diagram
  may be regarded as diagnostic of oscillations driven by gravothermal
  effects \citep{mak96}.  The three main loops, which are
  distinguished by different line types for clarity, correspond to the
  last three complete oscillations in
  Fig.~\ref{fig:gravo}.}\label{fig:gravo1}
\end{figure}

Now we consider the onset of gravothermal oscillations in our
simulations.  While gravothermal oscillations would not be expected to
occur in models as small as that shown in Fig.~\ref{fig:2}
\citep{goodman87}, for $N=16384$ deep gravothermal oscillations are
observed in single-component $N$-body systems (cf. \citealt{mak96}).
Indeed we observe gravothermal oscillations in our run with $N=
16384$.  However their onset occurs later than in the simulations of
Gao et al. (see Figs.~\ref{fig:gravo}, \ref{fig:gravo1}), though again
we do not know how the time of onset should vary with $N$; this
difference may or may not be significant.

Further evidence, though it is {contradictory}, is provided by the results
of other authors (Table \ref{tab:tb2}).  By comparison of their two
papers, \citet{gie03} concluded that the time of onset of gravothermal
oscillations is greatly affected by the treatment of binary
interactions.  The treatment of \citet{fre03} more closely resembles
that used by Giersz \& Spurzem in their earlier paper, {but they
  found (\citet{fre05}) that the onset of gravothermal oscillations is {\sl delayed} when binary
  interactions are computed directly.}

\begin{table}
\caption{Comparative results on the core radius and gravothermal
  oscillations for simulations with $f=10~\%$}
\label{tab:tb2}
\begin{center}
\begin{tabular}{lll}
Source	&$\delta\log_{10}r_c$	&$t_{gto}$\\
\hline
This paper	&$0.65$	&$130$\\
\citet{gie00}	&$0.2$	&$160$\\
\citet{gie03}	&$0.42$	&$20$	\\
\citet{gao91}	&$1$	&$50$	\\
\citet{fre03}	&$0.48$	&$70$	\\
{\citet{fre05}}	&$1$	&$\gtorder125$	\\
\hline
\end{tabular}
\end{center}
Notes: $\delta\log_{10}r_c$ is the change in the logarithm of $r_c$
from $t=0$ until the end of the initial contraction onto the
``deuterium main sequence''; $t_{gto}$ is the time (in units of
$t_{rh}(0)$) of onset of gravothermal oscillations.  Most of these
values have been estimated quite approximately from the figures in the
quoted papers.  {\sl}
\end{table}

\subsection{Total Mass in Singles and Binaries}

\begin{figure}
\resizebox{\hsize}{!}{\includegraphics{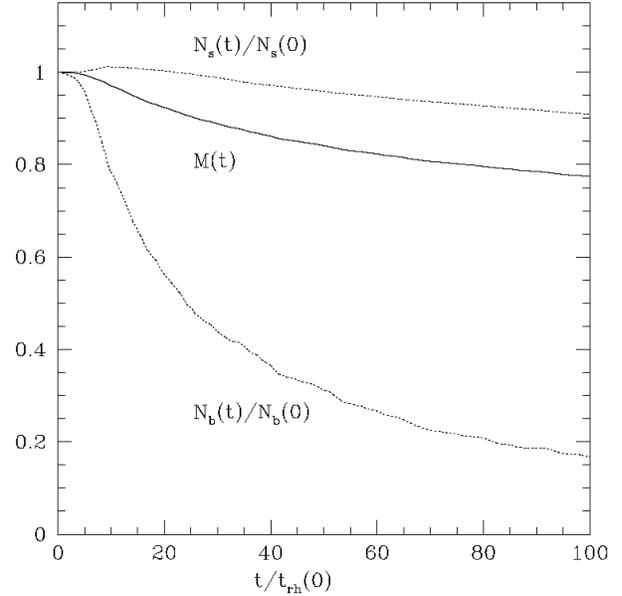}}
\caption{Time-dependence of the number of single (upper dotted line)
  and binary stars (lower dotted line) expressed as a fraction of the
  initial values.    The solid line is a graph of the total mass of the
  system.  The data refers to a simulation with $N=16384$ and $10\%$
  primordial binaries.  The unit of time is the initial half-mass
  relaxation time.}\label{fig:2a}
\end{figure}

The destruction of the binary population is shown in
Fig.~\ref{fig:2a}, which is comparable with Fig.~2a in Gao et al.  The
destruction of binaries takes place at a quite similar rate in the two
models, but escape of single stars is much larger in the $N$-body
simulations.  Indeed in the $N$-body models this somewhat diminishes the early {\sl increase} in the
numbers of single stars which is so prominent in the Fokker-Planck
model, and which is caused by
the destruction of binaries; nevertheless this feature is still
visible in our $N$-body results.  In the hybrid model of \citet{gie00} the number of single stars decreases monotonically, though the
residual mass in binaries at $t = 100 t_{rh}(0)$ is very similar to
the value in our Fig.~\ref{fig:2a}.

\begin{figure}
\resizebox{\hsize}{!}{\includegraphics{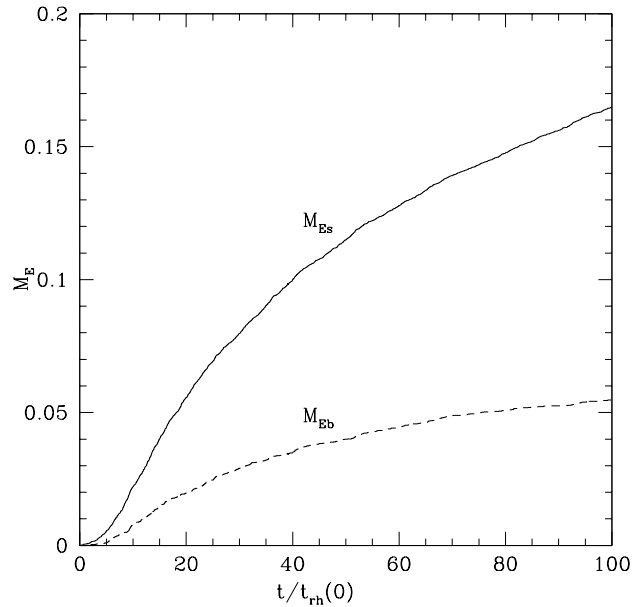}}
\caption{Evolution of the escaper mass in singles ($M_{Es}$, solid
line) and binaries ($M_{Eb}$, dotted line) for a simulation with
$N=16384$ and $10\%$ primordial binaries.}\label{fig:escapers}
\end{figure}

The cumulative mass lost by escape (separately in binary and single
stars) is shown in Fig.~\ref{fig:escapers}, which can be compared with
Fig.~27 in \citet{gie00}.  The mass of single escapers at time
$100\trh0$ is well matched, but we find that the mass of binary
escapers is about 50\% larger than in the hybrid model.

{In our model with $N=16384$, the decrease of the total mass by
  time $t = 100t_{rh}(0)$ is rather similar to that found by Fregeau
  et al (2003, Fig.4; 2005, Fig.2), but is about 
 50\% larger
in our model than theirs 
 by time $t = 200t_{rh}(0)$.}  We find that
the total potential and kinetic energy of the bound members varies
with $t$ in a manner very similar quantitatively to the result
depicted in their Fig.~2 {(\citet{fre03})}.

\subsection{Energy Distribution of Binaries}

\begin{figure}
\resizebox{\hsize}{!}{\includegraphics{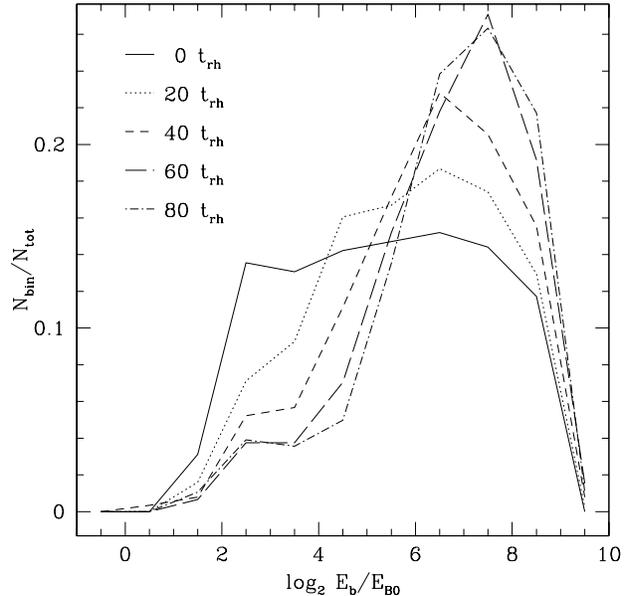}}
\caption{Evolution of the distribution of binary binding energy (given
as in NBODY6 in units of $E_{B0}=(3/2)kT$, where $T$ is the 
mean kinetic temperature of the initial Plummer model).  The
ordinate is the number of binaries per bin, which are of unit length
in $\log_2$. Curves have unit normalization, and are shown for
$t/t_{rh}(0) = 0, 20, 40, 60, 80$. The simulation has
  been performed with $8192$ particles and $10\%$ of
  binaries.}
\label{fig:2aN} 
\end{figure}

Fig.~\ref{fig:2aN} shows the evolution of the distribution of internal
binding energies of the binaries, and can be compared with Fig.~2b in
Gao et al. Both diagrams use a logarithmic energy scale (abscissa),
but with different definitions; for Gao et al. it is based on the
discrete energy levels they adopted, whereas in our case it is
determined by the binning in the output of NBODY6.  The range of
energies shown in the figure of Gao et al. (which is slightly smaller
than the range of energies in the initial conditions) corresponds in
our diagram to the range $2.1\ltorder\log_2(E_b/E_{B0}) \ltorder 8.5$,
where $E_{B0}$ is defined in the caption.

Harder binaries were not plotted by Gao et al., as they were treated
as being dynamically inert.  By the end of their run this group of
binaries accounted for about one third of the remaining binaries.  In
our $N$-body simulations, very few such extremely hard binaries are
found.  Simple theoretical ideas show that a binary is liable to be
ejected (in a binary-{\sl single} interaction) if $E_b \gtorder 15
m\vert\phi_c\vert$ \citep{goodman84}, where $\phi_c$ is the central
potential.  For a Plummer model this gives $E_b/E_{B0} \gtorder 100$,
which corresponds to a value of about $7$ in the units of the
horizontal axis of Fig.~\ref{fig:2aN}, in good agreement with our
results. The central potential deepens by a factor of only about 2
(Fig.~\ref{fig:phi0}).

We consider that the escape of very hard binaries in the $N$-body
simulations accounts for the main differences between our result
(Fig.~\ref{fig:2aN}) and that of Gao et al.  While we find that the
distribution of $E_b$ (at fixed time) declines at the highest binding
energies, in the results of Gao et al. the distribution of binding
energy is monotonically increasing.  Both kinds of simulation agree in
the depletion of less-hard pairs, though in Gao et al. the
depletion extends to somewhat harder pairs.  To quantify these effects
briefly, we remark that, for $N=16384$, the lower 10 percentile
increases by a factor of about 3.5, and the mean increases by a factor
of about 2.3, which is smaller because of the preferential destruction
of softer binaries.

\subsection{Spatial Distribution of Binaries}\label{sec:space_distn_binaries}

\begin{figure}
\resizebox{\hsize}{!}{\includegraphics{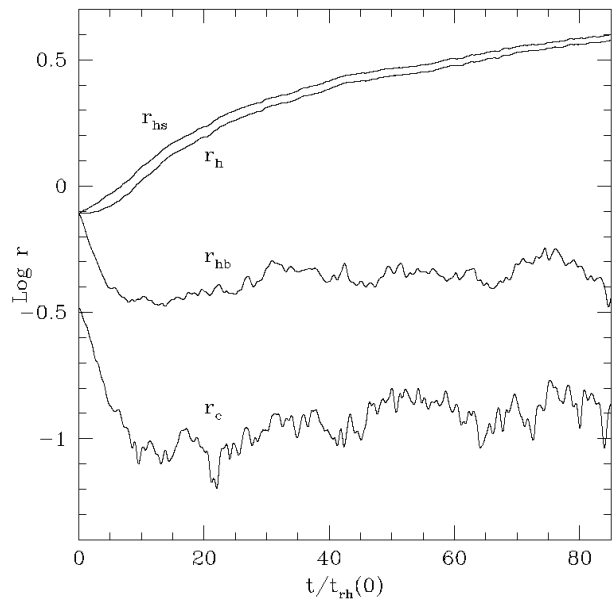}}
\caption{Dependence of half-mass radius (N-body units) for singles
  (upper curve), whole system (second curve) and binaries (central
  curve) on time (units of the initial half-mass relaxation time). The
  lowest curve is the core radius of the system. The simulation has
  been performed with $8192$ particles and $10\%$ of
  binaries.}\label{fig:3a}
\end{figure}

\begin{figure}
\resizebox{\hsize}{!}{\includegraphics{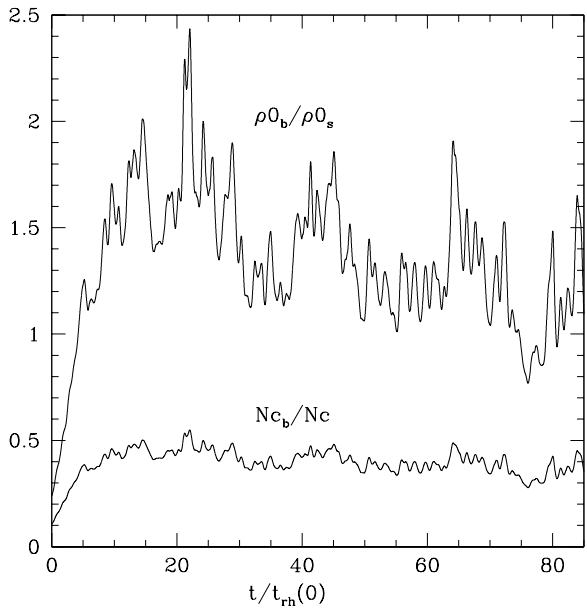}}
\caption{Binary fraction in the core, as a function of $t/t_{rh}(0)$.
The upper curve is the ratio of the mass density of binaries to the
mass density of single stars; the lower curve is the binary fraction
(corresponding to $f$, eq.(\ref{eq:f}), but restricted to the
core). The simulation has been performed with $8192$ particles and
$10\%$ of binaries.}\label{fig:3b}
\end{figure}

The evolution of the spatial distribution of binaries is indicated in
two ways in Figs.~\ref{fig:3a} and \ref{fig:3b}, which can be compared
with Fig.~3 in Gao et al.

Our results for the evolution of the half-mass radius of the single
and binary stars (Fig.~\ref{fig:3a}) are very similar to those of Gao et al. and \citet{gie00}, and rather at variance (at
least in later stages of the evolution) with those of \citet{fre03}; their Fig.~4 indicates the half-mass radius of the binaries
exceeding its initial value by about $70t_{rh}(0)$, and exceeding the
current half-mass radius  of the single stars after about
120$t_{rh}(0)$.  {In \citet{fre05} the evolution of the half-mass
radius is, if anything, still faster.}

Within the core the agreement with Gao et al. is poorer.  Our values
for the density ratio (in binaries and singles: Fig.~\ref{fig:3b}) are
considerably smaller than the values found by Gao et al.; those
authors find $\rho_{0,b}/\rho_{0,s}\simeq10$ towards the end of the
steady binary burning phase.  This quantity is $N$-dependent (see Sec.
\ref{sec:spatial_n}), but we do not consider this sufficient to
explain the disagreement.   (Extrapolation from our data suggests
$\rho_{0,b}/\rho_{0,s}\approx 3$ for the Gao et al. number of
particles).

From the dynamical point of view the early core contraction and
increase in $\rho_b/\rho_s$ can be seen as manifestations of mass
segregation.  \citet{spitzer69} showed that this could lead to
equipartition between two species (here binaries and single stars) if
the initial mass in the heavier species was sufficiently small.  His
criterion corresponds, in the present situation, to $f\ltorder0.03$.
Therefore all of the evolution we observe takes place in the regime
corresponding to the mass segregation instability.

\begin{figure}
\resizebox{\hsize}{!}{\includegraphics{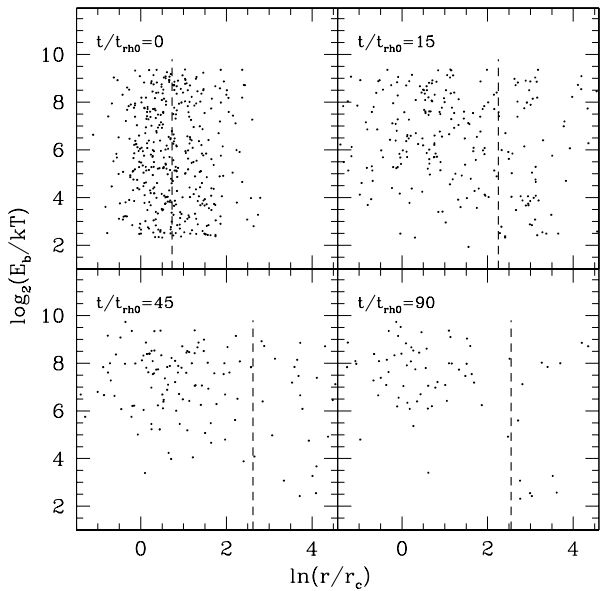}}

\caption{Evolution of the distribution of binaries in space of radius
and binding energy.  The radius is given with respect to the
instantaneous core radius; the energy in units of $kT$.  The panels
refer to $t/t_{rh}(0) = 0, 15, 45, 90$. The dashed line is the
position of the half mass radius. The simulation has been performed
with $4096$ particles and $10\%$ binaries. Comparison with similar
data in \citet{gie00} is given in the text. 
}\label{fig:r_eb}
\end{figure}

It is worth considering whether the topics discussed in this
subsection and the last are interconnected, i.e. whether the
distribution of binding energies is different at different
radii. Fig.~24 of \citet{gie00} implies that the removal of soft pairs
mainly affects binaries in the core.  They also find a gap in the
numbers of binaries at intermediate radii, which was predicted by
\citet{hmr92}. These trends seem to be confirmed in our simulations
(see Fig.~\ref{fig:r_eb} bottom right panel), but the number of
surviving binaries is too low to draw firm conclusions.

\subsection{Lagrangian Radii}

\begin{figure}
\resizebox{\hsize}{!}{\includegraphics{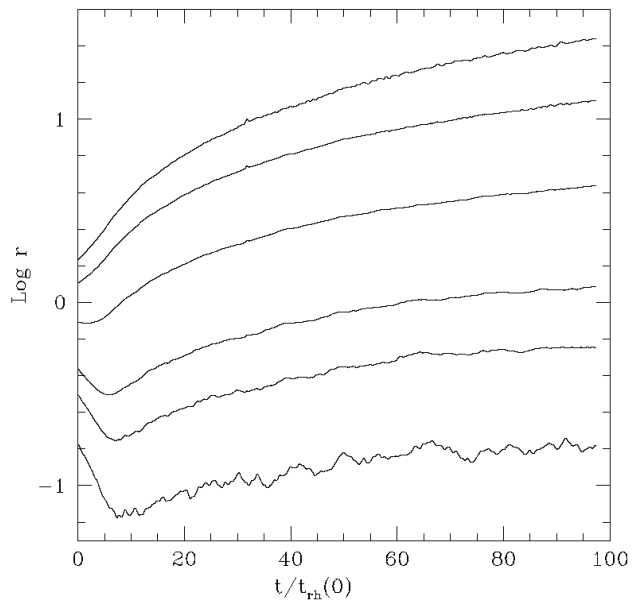}}
\caption{Lagrangian radii (enclosing $2,10,20,50,75,85 \%$ of the
  total mass), for the simulation with $N=16384$ and $10 \%$ of
  primordial binaries. }\label{fig:5}
\end{figure}

We have already discussed the evolution of the half-mass radius.
The evolution of other Lagrangian radii is displayed in
Fig.~\ref{fig:5}, which can be compared with Fig.~5 in Gao et al., except
that the choice of mass fractions differs at some points from theirs.
The differences in the Lagrangian radii are at the level of quantitative detail.  For example,
we find that the 10\% Lagrangian radius grows by about 0.1 dex up to
$t=50t_{rh}(0)$, whereas Gao et al. find an increase by about 0.3 dex.

\subsection{Structural Parameters}

\begin{figure}
\resizebox{\hsize}{!}{\includegraphics{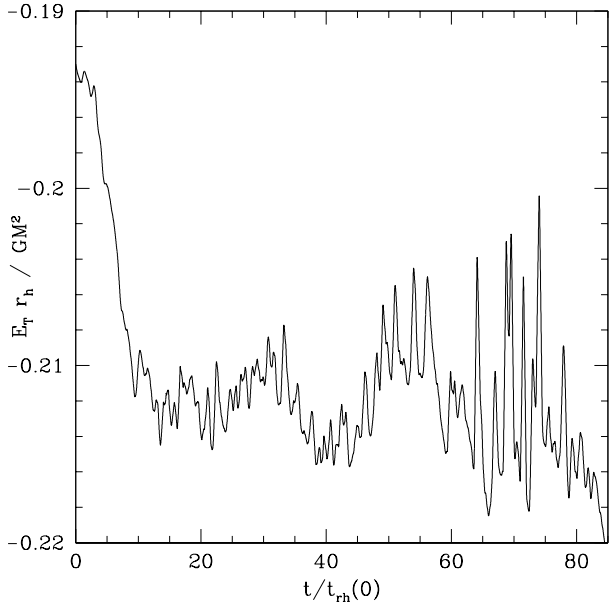}}
\caption{Structural parameter for the star clusters $E_{\rm T} r_h / G M^2
  $, where $E_{\rm T}$ is defined in Sec. \ref{sec:simulations}. To be
  compared with Fig 6 in Gao et al (though they plot separate curves
for $M\phi_0/E_{\rm T}$ and $r_hE_{\rm T}/(GM^2)$). The simulation has
  been performed with $8192$ particles and $10\%$ of
  binaries.}\label{fig:scaled_e}
\end{figure}

Gao et al. point out that much of the evolution of their model can be
described approximately as homological.  Part of the evidence for this
is their finding that some dimensionless measures of their model vary
little over the evolution, even though individual quantities, such as
the half-mass radius, may change by large factors.  We find, as do Gao
et al., that the structural parameter $E_{\rm T} r_h / G M^2 $ is
almost constant (Fig.~\ref{fig:scaled_e}; note the magnified vertical
scale).  

The value of
$-0.2$ is often adopted for this scaled energy  in simple models of bound
stellar systems (\citealt{spitzer87}, equation~(1-10)).

\begin{figure}
\resizebox{\hsize}{!}{\includegraphics{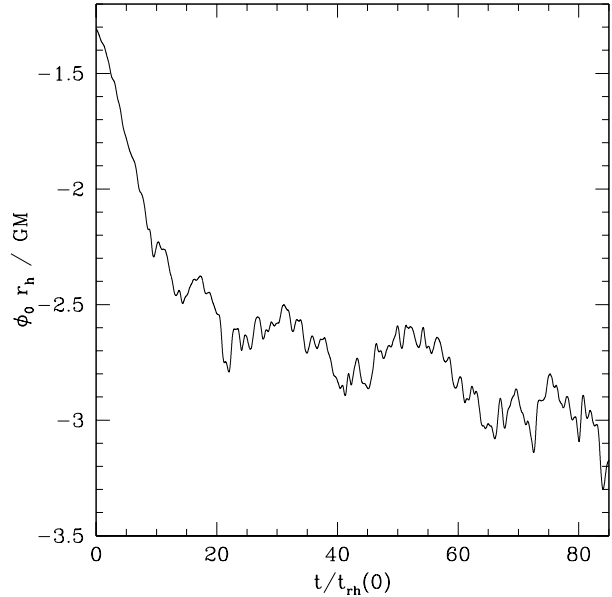}}
\caption{Structural parameter for the star clusters $\phi_0 r_h /
  GM$. To be compared with Fig 6 in Gao et al. The simulation has been
  performed with $8192$ particles and $10\%$ of
  binaries.}\label{fig:phi0}
\end{figure}

Another important quantity is the central potential, a scaled form of
which is presented in Fig.~ \ref{fig:phi0}.  Though our scaling differs from
that adopted by Gao et al., the fractional change between the initial value
and the value in the post-collapse phase is comparable.

\section{Dependence on N and binary fraction}\label{sec:ndependence}

\subsection{Core evolution}

\subsubsection{The time of core collapse}

\begin{figure}
\resizebox{\hsize}{!}{\includegraphics{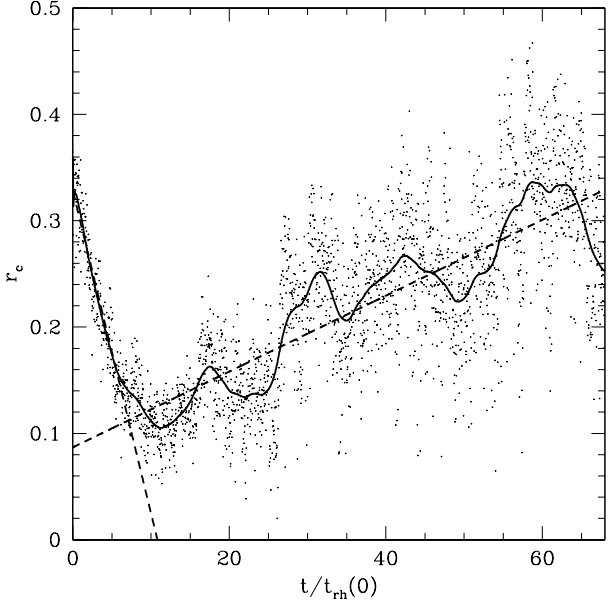}}
\caption{Determination of the time of core collapse. Smoothed and
  unsmoothed core radius for the model with $N=2048$ and $10\%$
  binaries.  The two linear graphs are explained in the text and are
  used to determine the time for core collapse.
  }\label{fig:findtcc}
\end{figure}

Determination of the time of core collapse requires first a careful
operational definition of this term, which is much less clear in these
models than it is in the absence of primordial binaries.  If $N$ is
large enough, when $f = 0$ there is a deep and sharp core collapse
(Fig.~\ref{fig:1}), and the determination of the time of core collapse
is relatively unambiguous.  It does, however, vary from one simulation
to another \citep{sa96}.  In the presence of primordial binaries the
problem is illustrated in Fig.~\ref{fig:findtcc}.  The initial
collapse of the core is almost linear, and we may determine the point
at which this ends ($t_{cc}$) by finding the intersection of two
linear fits to the data, for $t>t_{cc}$ and $t<t_{cc}$ respectively.
This has been done iteratively.  Alternatively, one may attempt to
determine the end of core collapse by finding the minimum of $r_c$.
This requires smoothing of the data, and in general the result will
depend on the smoothing interval.

Both methods present advantages and disadvantages. The linear fitting
method is very robust for the determination of the first line (i.e.
for $t<t_{cc}$), which plays the major role for setting the
\emph{time} of core collapse. In fact this first line is much steeper
than the second (that fits the core radius in the phase of post
collapse). This second fit is more sensitive to the noise and to the
extent of the fitting window (and thus in particular to the end time
of the simulation), so that the uncertainty associated with the
determination of the core \emph{radius} at the time of core collapse,
defined as the intersection of the two linear fits, is larger than the
uncertainty in the time. It may be also argued that, with this
definition, the core radius determination is somewhat artificial,
since it does not correspond to a realized (or averaged) value in the
simulation.  In fact, at the time of core collapse the core radius in
the simulation is bigger than the intersection of the two fits (see
Fig.~\ref{fig:findtcc}).

The second method, i.e. the determination of the time of core collapse
as the time at which the minimum value of $r_c$ is attained, may be
significantly affected by the intrinsic random fluctuations in the
measure, due to its sensitivity to the local details of the $r_c(t)$
curve. After some experimentations, we found that a good choice for
minimizing the errors is given by adopting a smoothing window with a
width of $2-3~t_{rh}(0)$ for $N \gtrsim 2048$ and slightly bigger
($3-4~t_{rh}(0)$) for $N \lesssim 2048$.  We note, however, that the
local nature of this estimate can play also a welcome role, since to
determine the time of core collapse, and the associated core radius,
the simulations may be stopped just shortly after core collapse. In
contrast, the double linear method requires data for several tens of
relaxation times during the postcollapse phase.

To evaluate the relative performance of the two estimators we have
applied both methods to a large number of simulations (a major
subsample of the simulations with $20\%$ primordial binaries; see
Table~1). As can be guessed from Fig.~\ref{fig:findtcc}, we found a
systematic bias of approximately $3 t_{rh}(0)$ between the two methods
in the measure of the time of core collapse, with the linear fit
giving smaller times. The numerical experiments also confirm that the
variance of the measure is marginally smaller ($30\%$ less) for the
linear fit method; the results are only very weakly N-dependent. The
values for the core radius at core collapse are, on the other side,
similar for the two methods.

In what follows we present results from the second method (minimum of
$r_c$) for determining $t_{cc}$ and $r_{cc}$, and the possible
limitations of its significance need to be borne in mind. The main
reason for choosing this procedure is the fact that this has allowed
us to consider as valid simulations all those that reached at least
$10~t_{rh}(0)$ after core collapse. In fact, while formally all the
runs should have lasted for at least $100 t_{rh}(0)$, some of them
were stopped prematurely due to numerical difficulties. Given the
large number of experiments (see Table 1), it was impractical for us
to manually restart all the runs that required assistance. For each
entry in Table 1 we have, however, at least one completed simulation,
and on average the completeness of the sample is above $75\%$.

\begin{figure}
\resizebox{\hsize}{!}{\includegraphics{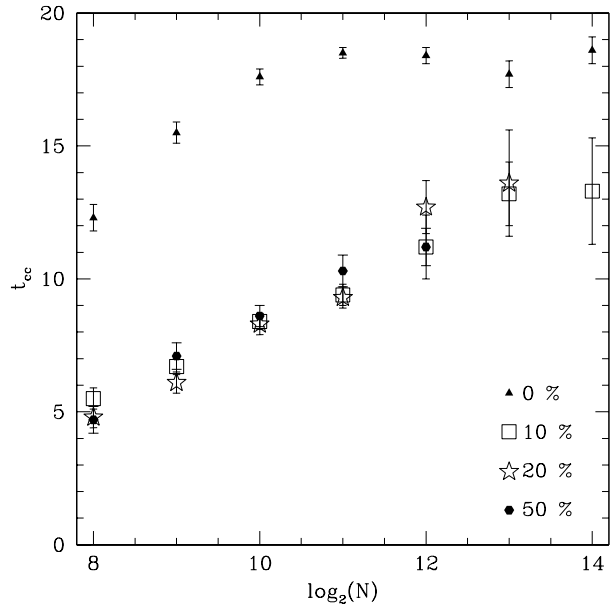}}
\caption{Time for core collapse for the set of simulations with $0\%$
to $50\%$ primordial binaries.  The filled triangles correspond to $f
= 0$.}\label{fig:tcc}
\end{figure}

Our results are shown in Fig.~\ref{fig:tcc}.  From this figure it may
be concluded that, at a given value of $N$, the time of core collapse
is essentially independent of the initial binary fraction $f$, if
$f\gtorder 10$\%, and smaller than the collapse time of a model
without primordial binaries by a factor of roughly $2$. This compares
well with the expectation based on a two components model with single
stars only (and mass ratio 1:2) studied by \citet{ina84}, that would
predict a relative ratio of the order $1.6 \pm 0.2$.  While this lack
of dependence on $f$ may seem surprising, it should be observed that,
at late times in core collapse, the binary fraction in the core has
increased considerably (Fig.~\ref{fig:nbns}), to 50\% by number (or
more) even if $f = 10$\% initially.  Therefore the core is practically
saturated with binaries, approximately independent of the initial
value of $f$.  Under these circumstances it is not surprising that the
behavior of the core varies little with $f$. Further evidence of this
``saturation'' effect is visible below in Figs.\ref{fig:rcrh} and
\ref{fig:rc_VC}.

\label{sec:gto}

\citet{fre03} give a plot showing that the ``core-collapse time''
exhibits a roughly linear increase with $f$.  This is quite at odds
with our result, but the issue is a semantic one.  \citet{fre03} do
not call the initial contraction ``core collapse'',
preferring to reserve this term for the deep collapses akin to
gravothermal oscillations; these occur much later, when the binaries
are almost fully depleted.  Thus the time they discuss might
correspond to the deeper collapses in Fig.~\ref{fig:gravo} after about
$t \sim 150t_{rh}(0)$, as discussed in Sec.~\ref{sec:rcrh}, whereas
our figure refers to the end of the initial phase of contraction at
about $t = 13t_{rh}(0)$.

\subsubsection{Core radius in the phase of steady binary burning}\label{sec:rc}

\begin{figure}
\resizebox{\hsize}{!}{\includegraphics{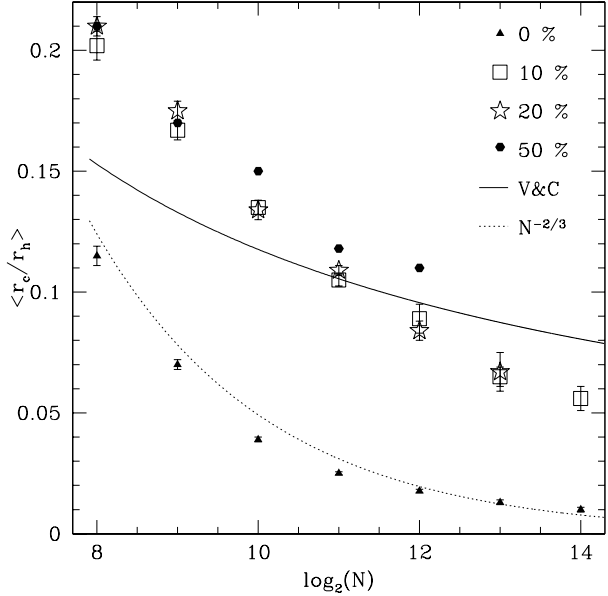}}
\caption{Core to half mass radius averaged over a period of $10
t_{rh0}$ after core collapse and over different realizations (see
Table~\ref{tab:tb1}) for the set of simulations with $0\%$ to $50\%$
primordial binaries. The points associated with runs with primordial
binaries are compared to the Vesperini \& Chernoff model with $\gamma
= 10$, $v_c/v_h = \sqrt{2}$ and $\phi_b = 0.4$. The points associated
with single-stars runs are compared to a $N^{-2/3}$ power law
\citep{heg03}, Box~28.1\label{fig:rcrh}\label{fig:comparison}}
\end{figure}

The information from our simulations on this topic is summarised in
Fig.~\ref{fig:rcrh}.  We notice again a saturation effect for initial
binary fractions $f\gtorder 0.1$, and a significant variation with $N$.

To explore these dependences further, we now compare the ratio of the
core to half mass radius in the post-collapse phase with the
theoretical estimate given by \citet{vc94}.  By balancing the energy
production in the core with the rate of expansion at the half mass
radius, they get the following estimate for this ratio (their eq.~9):
\begin{equation}
\frac{r_c}{r_h} = \frac{\alpha}{\log_{10}(0.4N)}
\frac{\phi_b (1-\phi_b) \mu_{bs} + \phi_b^2 \mu_{bb}}{(1+\phi_b)^4},\label{eq:vc}
\end{equation}
where the various quantities are defined as follows.  The quantity
$\alpha$ is a parameter depending on $\gamma$ and $v_c/v_h$, where $\gamma$ is the ratio of the expansion timescale $r_h/\dot r_h$
to $t_{rh}$, and $v_c$ and $v_h$ are the velocity dispersion in the core
and at the half-mass radius, respectively;  we have assumed typical
values for these parameters: $\gamma \approx 10$ and $v_c/v_h \approx
\sqrt{2}$ \citep{heg92,goo89}. The quantity $\phi_b$ is the binary fraction in the
core defined as
$$
\phi_b = \frac{N_b}{N_s+N_b}.
$$ 
Finally, $\mu_{bs}$ and $\mu_{bb}$ are coefficients for the efficiency of
binary-single and binary-binary burning and depend on the distribution
of binding energy of the binaries. 
These coefficients have been computed by a numerical implementation of
Equation~9 in \citet{vc94}, which has been checked against the
results given by those authors. For simplicity, to compute the
interaction cross-sections we have assumed a flat distribution in
$\log(E_b)$ between $12$ and $300$ kT, where the upper and lower limits
have been estimated from the measured binding energy distribution (see
Fig.~\ref{fig:2aN}).  

For a fixed initial binary fraction $f = 0.1$ this model is compared
with the results of our simulation in Fig.~\ref{fig:comparison}
($\alpha$ is assumed to be constant). Clearly the $N$-dependence is
different. This result appears indeed rather puzzling.  In fact it is
unlikely that differences in the current fraction of binaries in the
core, $\phi_b$, are responsible, because of the saturation effect. The
$\mu_{bs}$ and $\mu_{bb}$ coefficients are also $N$-independent, since
we have checked that the distribution of binary binding energy, that
is the only relevant input, evolves remarkably similarly in
simulations with $N$ from $512$ to $16384$. In the worst case, our
simplified approach (i.e. assuming a uniform distribution in
$\log(E_b)$ for computing the analytical model) thus introduces only a
\emph{constant} multiplicative bias in eq.~\ref{eq:vc}. In addition,
this bias should be limited, since Figs.  5 and 6 in \citet{vc94}
suggest that the core radius is not particularly sensitive to the
upper and lower limits of this distribution, at least within the
limits of variation we observe (Fig.~\ref{fig:2aN}).  Other factors
which may influence the efficiency of binary heating are the total
mass and central potential, but $M$ varies with $N$ by at most about
$2\%$ up to $t/t_{rh}(0) = 100$ and $\phi_0$ does not show any
systematic trend with $N$ (with an average value, after the core
collapse, of $\approx 1.6$ in NBODY6 units for our runs with
$f=10\%$).

An interesting issue, raised by the referee, is the time scale on
which binaries release energy.  The time scale for close interactions
between a binary of semi-major axis $a$ and single stars of space
density $n$ may be estimated from the cross
sections of \citet{1984ApJS...55..301H} as approximately
$t_{bs}\simeq v_{inf}/(24nGma)$, where $v_{inf}$ is the relative velocity of a
binary and third body when far apart, 
and $m$ is the stellar mass.  Comparing with the local relaxation time $t_r$
(\citet{spitzer87}), we find that $t_{bs}/t_r \simeq
0.9Gm\ln\Lambda/(av^2)$, where $v$ is the root mean square
three-dimensional velocity of single stars and $\ln\Lambda$ is the
Coulomb logarithm.  Since this depends on $N$ there is an
$N$-dependent regime of binaries which are effectively inert on the
time-scale of a given simulation, and this potentially introduces an
$N$-dependence in the efficiency of binary interactions.  On the other hand the time scale for
interactions involving our hardest primordial binaries is
approximately $240 t_r$.  Since the central relaxation time is much
smaller than the half-mass relaxation time, we conclude that all
primordial binaries have enough time to interact within the time scale
we have considered (up to $100t_{rh}(0)$).

We have also measured the value for $\gamma$ in the simulations,
confirming, immediately after core collapse, the values reported by
\citet{heg92}, i.e. $\gamma \approx 12$, and observing basically no
$N$-dependence.  In this regard we note that Gao et al. measured
$\gamma = 7$.  From our set of simulations this value appears to be
unexpectedly low, even when account is taken of the fact that their
model is isotropic (cf. Fig.2 in Takahashi 1996).  Though we are unable to
explain the discrepancy between the values of $\gamma$ it does offer
an alternative explanation of the fact that our values of $r_c/r_h$
appear to disagree with those of Gao et al (cf. our Sec.4.1).
According to the theory of Vesperini \& Chernoff a change in the value
of $\gamma$ leads to a change in the value of $r_c/r_h$ consistent
quantitatively with what is found.

Finally, the differences in the value of $r_c/r_h$ at core collapse
could be due to variations of the central velocity dispersion $v_c$,
which could have a significant impact, as $\alpha \propto
(v_c/v_h)^3$. From our simulations we do not observe a significant
$N$-dependence for this quantity, which appears to be close to the
values quoted by \citet{goo89}, $(v_c/v_h)^2 \approx 2$. To be sure a
limited $N$-dependence is present, but this goes in the opposite
direction with respect to the one required to obtain a better match in
Fig.~\ref{fig:rcrh}, so that if this trend is taken into account the
comparison with the analytical estimate for $r_c$ becomes worse: for
$10\%$ primordial binaries we measure $(v_c/v_h)^2 = 1.88 \pm 0.03$
with $N=256$, $(v_c/v_h)^2 = 1.95 \pm 0.02$ for $N=1024$ and
$(v_c/v_h)^2 = 2.07 \pm 0.02 $ for $N=4096$. In contrast Gao et
al. report a lower value, $(v_c/v_h)^2 \approx 1.8$, so that this low
central velocity dispersion could account for the smaller core in
their Monte Carlo run, although it is not clear what is the origin of
such a low velocity dispersion. Further studies would be required to
clarify this point.

To summarize, from the data obtained in our simulations each term that
contributes to the \citet{vc94} model cannot explain why $r_c/r_h$
exhibits a steeper variation in $N$ than expected. Since the number of
particles for the lower $N$ tail of the curve in Fig.~\ref{fig:rcrh} is
rather modest, one possible explanation could be due to the importance
of granularity effects so that a mean field approximation is not fully
justified (see also the concluding discussion in Sec.~\ref{sec:con}).

\begin{figure}
\resizebox{\hsize}{!}{\includegraphics{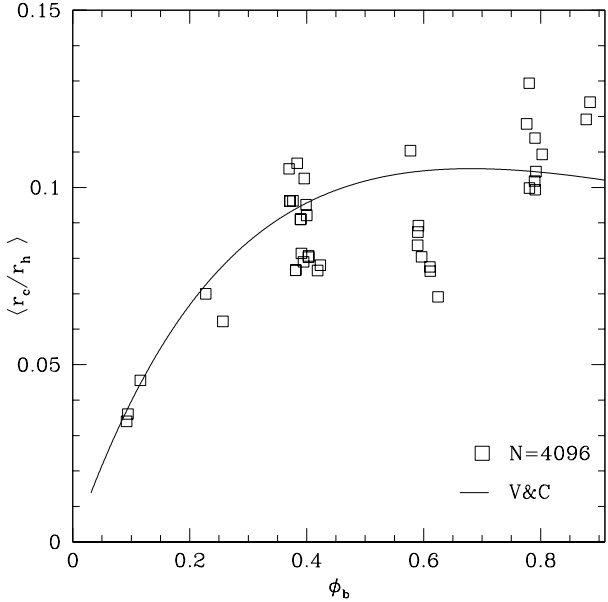}}
\caption{Core to half mass radius averaged for $10 t_{rh0}$ after the
  core collapse for the set of simulations with $N=4096$ compared with
  the Vesperini \& Chernoff model. For this model (Eq.~\ref{eq:vc}) we have adopted
  $\gamma = 9.4$, $v_c/v_h = \sqrt{2}$, distribution in the binary
  binding energy flat in log scale from $12$ to $300 kT$. These values
  imply $\mu_{bs}=2.841$ and $\mu_{bb}=0.774$. 
}\label{fig:rc_VC}
\end{figure}

At fixed $N$ a comparison with eq.(\ref{eq:vc}) 
is shown in
Fig.~\ref{fig:rc_VC}. Here $\alpha$ is kept as a free parameter. In
this case, there is satisfactory quantitative agreement within the
scatter of different simulations at each value of $f$, except possibly
for the highest values.  (Note that we here use the {\sl current}
value of the binary fraction, $\phi_b$, averaged over the same time
interval, and not the initial value.  The two groups on the right,
i.e. at $\phi_b\sim0.8$ and $0.9$ correspond to initial values of $f =
0.5$ and $1$, respectively.)  For $N = 2048$ (not shown) we obtain
comparable agreement for all $f$, except for an indication that the
theory predicts slightly too small a core radius for the very highest
and the very lowest values of $f$.  It must be borne in mind, however,
that the core radius in a multi-component system is not a well defined
concept. Our data for $N = 2048$ do not exhibit the slightly puzzling
clumping at $\phi_b \simeq 0.4$ in Fig.~\ref{fig:rc_VC}, or the
systematic offset at $\phi_b \simeq 0.6$.  Our dataset for $N=2048$ is
considerably larger, and we conclude that these features in
Fig.~\ref{fig:rc_VC} are results of the more modest sample size.

\subsection{The binary and single populations}

\subsubsection{Total mass in singles and binaries}

\begin{figure}
\resizebox{\hsize}{!}{\includegraphics{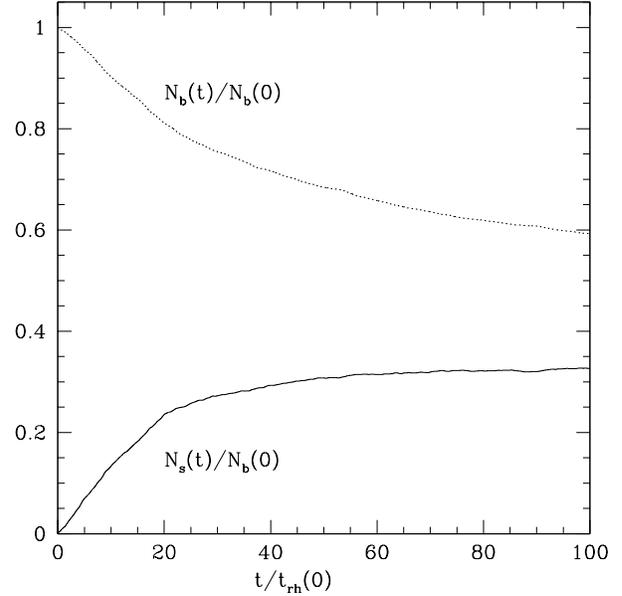}}
\caption{Dependence of number of single and binary stars (relative to
the initial values) on time (units of the initial half-mass relaxation
time).  Simulation starting with $4096$ primordial binaries and no
singles.}\label{fig:2a100}
\end{figure}

When the initial binary fraction is 10\%, the binaries are mostly
destroyed in the first 100$t_{rh}(0)$.  By the end of the simulation
they account for only about 5\% of the mass (Fig.~\ref{fig:2a}).  The situation is quite
different for a simulation which contains only binaries initially
(Fig.~\ref{fig:2a100}).  By the same point in this simulation the
binaries account for about 80\% of the remaining mass. 

\subsubsection{Spatial Distribution of Binaries}\label{sec:spatial_n}

\begin{figure}
  \resizebox{\hsize}{!}{\includegraphics{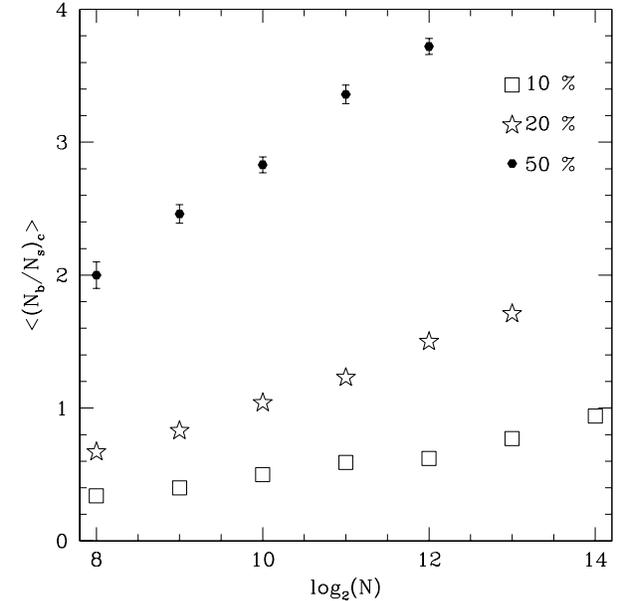}}
\caption{Ratio of the number of binaries to singles in the core
  averaged for $10 t_{rh0}$ after the core collapse for the set of
  simulations with 10\%, 20\% \& 50 \% primordial binaries.}\label{fig:nbns}
\end{figure}

The concentration of binaries in the core, after the end of core
collapse, is presented as  a function of $N$ and $f$ in
Fig.~\ref{fig:nbns}.  Some comparison is possible with the results of
Gao et al. at fixed $N$ (their Fig.~3b), who found that the ratio
increased by $0.3$ dex from $f = 0.1$ to $f = 0.2$.  An increase by
this factor is
consistent with what we find for all $N$ that we studied, but we
repeat that the actual values we find are much smaller than those
displayed by Gao et al. (Sec.~\ref{sec:space_distn_binaries}).

\begin{figure}
\resizebox{\hsize}{!}{\includegraphics{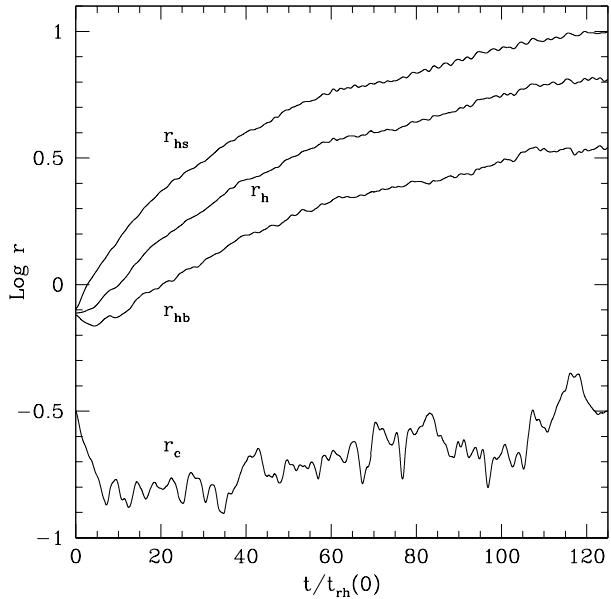}}
\caption{Dependence of half-mass radius (N-body units) for singles
  (upper curve), whole system (second curve) and binaries (central
  curve) on time (units of the initial half-mass relaxation time). The
  lowest curve is the core radius of the system. The simulation has
  been performed with $4096$ particles and $50\%$ of binaries. Unlike
  Fig.~\ref{fig:3a}, here, due to the higher ratio of primordial
  binaries, the half mass radius for binaries expands steadily after a
  short initial contraction.}\label{fig:3a_50}
\end{figure}

While Fig.~\ref{fig:nbns} confirms that the concentration of binaries
in the core increases with the original binary fraction, $f$, their
overall spatial distribution differs less from that of the single
stars, as we consider models with larger $f$.  This result is
illustrated in Fig.~\ref{fig:3a_50}, which should be compared with
Fig.~\ref{fig:3a} (see caption to Fig.~\ref{fig:3a_50}).  In this
respect our result differs from that of \citet{fre03}: as we have
already remarked (Sec.~\ref{sec:space_distn_binaries}), when $f=10$\%
they find that the half-mass radius of binaries exceeds that of
singles for $t/t_{rh}(0)\gtorder 120$. This does not happen in their
run with 20\% binaries initially (their Fig.~5), where $r_{h,b} <
r_{s,b}$ at all times up to the end of their run, at
$t/t_{rh}(0)\simeq 600$.

\section{Conclusion}\label{sec:con}

This paper has had several objectives.  Our first has been to provide
evidence from $N$-body simulations of the evolution of highly
idealised stellar systems containing an initial population of
binaries.  The Fokker-Planck models of Gao et al. have become an
important touchstone for subsequent modelling, but until now it has
not been clear whether any disagreement with their results is due to
the simplifying assumptions they made by adopting an approximate
method.  What we have tried to do with our simulations is to establish
what actually happens, so that subequent research can rest on fewer
assumptions and approximations. In this respect our work can be
considered as a stepping stone to improve the physical understanding
of the results from more complex and realistic numerical simulations,
such as those recently performed by \citet{por04}.

In line with this objective we have presented detailed results
following closely the exposition of Gao et al., which covers the
evolution of the structure of the system (core and half-mass radius,
and structural parameters such as the scaled central potential), and
the evolution of the binaries:  their total numbers, their abundance
in the core, their energies, and so on.

Our second objective has been to begin a discussion of the differences
between the various models which have appeared in the literature so
far, with particular reference to those of Gao et al., \citet{gie00}
and \citet{fre03,fre05}.  Here our discussion has covered several of the
above issues, and also the gravothermal oscillations which have been
observed in {almost} all simulations, but with somewhat contradictory results
for some characteristics, such as the time of onset of the
oscillations.  We have also compared our results with some of the
small amount of quantitative theoretical predictions in the
literature, mainly those of \citet{vc94}. With respect to this
analytical estimate for the core to half mass radius in the
post-collapse phase we find good quantitative agreement (at any {\sl
fixed} value of $N$) for the dependence on the initial binary
ratio. On the other hand the observed $N$-dependence of this quantity
is much steeper than one would expect on the basis of the theoretical
arguments given by \citet{vc94}. This could be simply a low-$N$
effect, but possibly some of the assumptions underlying the model
could be inaccurate.   

This issue is relevant to the comparison of the core radius between
our models and Fokker-Planck or Monte Carlo models.  With the present
data we cannot give a clear answer to this point, since either
extrapolation or direct simulations with $N \gtrsim 10^5$ would be
required. A different way of gaining some insight into this problem
could be to run some comparison simulations with Monte Carlo methods
with $N$ in the range considered in this paper. If our results would
be reproduced, they could be used to indirectly validate Monte Carlo
simulations with $N$ high enough to answer this open question.

Our third objective has been to attempt to add some new insight and
information on the nature of the collisional evolution of stellar
systems with an initial population of binaries.  In particular we have
considered the time of core collapse (the definition of which we have
attempted to clarify).  For the period of steady binary burning which
follows this initial collapse,  we have  determined the manner in which
the core properties (core radius and binary fraction) depend on the
initial binary fraction and particle number.

It is not clear which of these conclusions are applicable, even
qualitatively, to real star clusters, with an initial mass function, a
largely unknown initial distribution of binary parameters, a tidal
field, and stellar and binary evolution.  For example,
\citet{ivanova05} have argued (on the basis of a simplified treatment
of the core) that the binary fraction in the core of a dense cluster
may be as little as 5\%, even if $f=1$ initially.  Now \citet{fre03}
find (with a Monte Carlo code) that a system in a tidal field with
$f\gtorder0.1$ initially disrupts after at most 45$t_{rh}(0)$.  If we
were to guess from our Fig.~\ref{fig:2a100}, we would conclude that
the binary fraction at this time would actually exceed 70\%.  Of
course this argument ignores the essential physical issues of stellar
evolution, collisions, etc., which underly the conclusion of
\citet{ivanova05}. On the other hand, {\sl all} these models neglect
factors which may be essential.  Our next paper on this topic will
provide detailed information on one of these {which we have ignored so
far} -- the influence of the
tide.

\section{Acknowledgments}

We are indebted to Sverre Aarseth for the provision of NBODY6 and we
are grateful to Enrico Vesperini for helpful discussions and
suggestions. The input from the referee, David Chernoff, was both helpful and
interesting.  DCH and MT thank the Institute for Advanced Study for
its hospitality while part of this work was being carried out.  PH and
MT similarly thank the School of Mathematics at Edinburgh University for its
hospitality during later stages of this project. MT thanks
Prof. Mineshige for his kind hospitality at the Yukawa Institute at
Kyoto University, through the Grant-in-Aid 14079205 of the Ministry of
Education, Culture, Sports, Science and Technology,
Japan. PH thanks
Prof. Ninomiya for his kind hospitality at the Yukawa Institute at
Kyoto University, through the Grant-in-Aid for Scientific Research on
Priority Areas, number 763, "Dynamics of Strings and Fields", from the
Ministry of Education, Culture, Sports, Science and Technology,
Japan. 
The numerical simulations have been performed on the Condor
Cluster of the Institute for Advanced Study.


\bsp

\label{lastpage}

\end{document}